# Four-Photon Greenberger–Horne–Zeilinger Entanglement via Collective Two-Photon Coherence in Doppler-Broadened Atoms

Jiho Park, Heonoh Kim, and Han Seb Moon,*

*Department of Physics, Pusan National University, Geumjeong-Gu, Busan 46241, Korea*

*hsmoon@pusan.ac.kr

Bright, entangled multiphoton sources based on atom–photon interactions are an essential requirement in the realization of several quantum information and quantum computation schemes based on photonic quantum systems. Here, we experimentally demonstrate a four-photon polarization-entangled Greenberger–Horne–Zeilinger (GHZ) state obtained from Doppler-broadened atomic ensembles of $^{87}$Rb atoms. Owing to collective two-photon coherence in the Doppler-broadened cascade-type atomic system, our setup enables the generation of robust four-photon GHZ states with a fidelity of 82.1% and a measured four-photon average coincidence rate of 0.58/s. We believe that the generation of such bright and stable multiphoton GHZ states from atomic media is an important step toward realizing photonic quantum computation and practical quantum networks based on atom–photon interactions.

The entanglement of multipartite systems is an important resource for quantum information processing applications such as quantum computing and quantum secret sharing [1-5]. In this context, since the Greenberger–Horne–Zeilinger (GHZ) state was first proposed in 1989 [6], the generation of GHZ states has been experimentally demonstrated in various physical systems composed of photons [7], atoms [8], and ions [9]. In particular, the multiphoton entangled state forms a crucial quantum resource for photonic quantum information processes such as linear optical quantum computation (LOQC) and quantum communications [10-12].

In this context, multiphoton polarization-entangled GHZ states generated via spontaneous parametric down-conversion (SPDC) with nonlinear crystals have been intensively studied and applied to quantum algorithms and quantum teleportation [13-17]. Recently, by combining several polarization-entangled photon pairs in multiphoton interferometers, researchers have experimentally generated a twelve-photon GHZ entangled state via the SPDC process in a nonlinear crystal [13]. Moreover, the generation of a frequency-degenerate four-photon entangled state with the use of silicon nanowires has also been reported [18]. However, to effectively realize photonic quantum information processing, interactions between the multiphoton GHZ state and atoms are required for applying quantum memory based on atomic ensembles. In addition, to extend entangled multiphoton generation with the use of more than two independent photon sources, spectral overlap and spectral narrowing between two independent photon sources are important issues hindering the realization of effective entangled-multiphoton sources with high fidelity and brightness. In this context, we note that the narrowband multiphoton entanglement generated from atomic ensembles can intrinsically fulfill the requirement of indistinguishability.

In the above context, multiphoton GHZ entangled states generated from atomic ensembles are preferable over those generated via SPDC because atomic-ensemble-based multiphoton GHZ states afford suitable resonant wavelengths and optical bandwidth for atom–photon interactions. In this regard, a recent study has reported on the experimental realization of the narrowband four-photon Greenberger–Horne–Zeilinger state in a single cold atomic ensemble [19]. While the method used in the study is effective to demonstrate multiphoton GHZ entangled states generated from atomic ensembles, it only yielded a low four-photon coincidence count rate, corresponding to 0.04 events/s in a basis of the GHZ state with a fidelity of 65%. In particular, we note that it is difficult to realize bright, high-fidelity entangled-multiphoton sources from atomic ensembles for application to practical quantum information processing.

In this Letter, for the first time, we experimentally demonstrate the four-photon polarization-entangled GHZ state with a high generation rate and a high fidelity via the collective two-photon coherence from warm atomic ensembles of $^{87}$Rb atoms. Furthermore, our apparatus is simple, and it can be continuously operated with low pumping powers of the sub-milliwatt order. For characterizing the four-photon GHZ state from two independent warm atomic ensembles, we measure the fourfold coincidence in the horizontal and vertical polarization (H/V) basis of a GHZ state, and we perform quantum-state tomography of the four-photon GHZ state. In addition, we observe the time-resolved Hong–Ou–Mandel (HOM) interference of the four-photon entanglement in the continuous wave (CW) mode; this is possible because the photon coherence time is longer than the time-resolution of the single-photon detectors. Thus, our works can aid in extending photonic quantum resources to quantum computation and quantum communication.

Figure 1(a) shows the energy-level diagram of the $5S_{1/2}$–$5P_{3/2}$–$5D_{5/2}$ transition of $^{87}$Rb atoms for photon-pair generation via the spontaneous four-wave mixing (SFWM) process [20]. Two-photon coherence is generated between the $5S_{1/2}$ and $5D_{5/2}$ states of $^{87}$Rb via two-photon resonance with pump and coupling lasers. Here, we note that the warm atomic ensemble in an atomic vapor cell has a Maxwell–Boltzmann velocity distribution; however, it can be coherently two-photon excited upon satisfying the Doppler-free two-photon resonant condition due to the counter-propagating configuration of the pump and coupling lasers. Furthermore, because the wavelength difference between both lasers is small, it is possible to obtain collective two-photon coherence in the Doppler-distributed moving atoms. Owing to the collective two-photon coherence, the probability amplitudes of the signal and idler photons can be coherently superposed throughout the atomic spatial and velocity distributions in the warm atomic ensemble [21]. The signal and idler photons generated in the phase-matched direction exhibit frequency–time, polarization, and Einstein–Podolsky–Rosen (EPR) entanglements [21-25]. To practically realize the high-quality entangled state, the optical frequencies of both lasers are far-detuned by ~1 GHz from resonance to reduce the number of non-correlated photons due to one-photon resonance.

The polarization-entangled photon pairs from a warm atomic vapor cell are generated based on Sagnac configuration, which uses a simple setup and is robust to ambient noise [24]. The atomic vapor cell used in our experiment was a cylinder with a length of 12.5 mm and a diameter of 25 mm. In our experiment, the pump and coupling laser powers were 0.15 mW and 1.5 mW, respectively, with 1.2-mm diameters. In the study, we prepared the Bell states of two independent polarization-entangled photon-pair sources (S1 and S2), respectively. To confirm the realization of polarization-entangled Bell states, with the polarizer angle of the idler photon fixed at 45°, we measured the coincidence counts as a function of the polarizer angle of the signal photons (polarizers P1 and P4). The raw visibilities of the interference fringes of the two polarization-entangled photon-pair sources were estimated to be ~93%.

Next, we describe the procedure for the generation of the four-photon GHZ state from two independent polarization-entangled SFWM sources. As shown in Fig. 1(b), to simultaneously operate the two entangled photon sources (S1 and S2), the pump and coupling lasers are orthogonally linearly polarized and split equally at the beam splitter (BS). In our experiment, we prepared polarization-entangled states $\left|\Phi^+\right\rangle_{12}$ and $\left|\Phi^+\right\rangle_{34}$ corresponding to S1 and S2, respectively. Input state $\left|\Psi^i\right\rangle_{1234}$ can be expressed as

$$\begin{aligned}\left|\Psi^i\right\rangle_{1234} &= \left|\Phi^+\right\rangle_{12} \otimes \left|\Phi^+\right\rangle_{34} \\ &= \frac{1}{\sqrt{2}}\left(\left|H\right\rangle_1\left|H\right\rangle_2 + \left|V\right\rangle_1\left|V\right\rangle_2\right) \\ &\quad \otimes \frac{1}{\sqrt{2}}\left(\left|H\right\rangle_3\left|H\right\rangle_4 + \left|V\right\rangle_3\left|V\right\rangle_4\right).\end{aligned} \quad (1)$$

To obtain the polarization-entangled GHZ state, both idler photons 2 and 3 are directed to the two inputs of a polarizing beam splitter (PBS). After passage through the PBS, we project the input state onto the four-photon polarization-entangled GHZ state, which can be described as

$$\left|G_4\right\rangle_{12'3'4} = \frac{1}{\sqrt{2}}\left(\left|H\right\rangle_1\left|H\right\rangle_{2'}\left|H\right\rangle_{3'}\left|H\right\rangle_4 + \left|V\right\rangle_1\left|V\right\rangle_{2'}\left|V\right\rangle_{3'}\left|V\right\rangle_4\right), \quad (2)$$

where 2′ and 3′ denote the output modes of the PBS. We can analyze the $\left|G_4\right\rangle_{12'3'4}$ state in the H/V basis according to the four polarizers (P1, P2, P3, and P4) used in the setup.

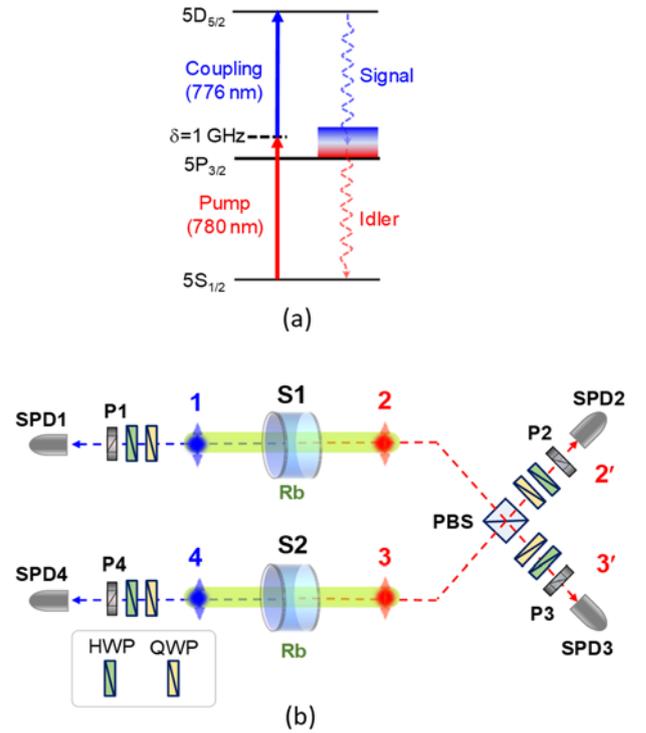

**Figure 1. Experimental configuration for the generation of the four-photon Greenberger–Horne–Zeilinger (GHZ) state from Rb atomic vapor.** (a) Configuration for correlated photon-pair generation via spontaneous four-wave mixing (SFWM) in a three-level cascade-type atomic system. (b) Schematic of the experimental setup for the measurement of the four-photon polarization-entangled GHZ state (P: polarizer; PBS: polarizing beam splitter; HWP: half-wave plate; QWP: quarter-wave plate; SPD: single-photon detector).

We consider two independent polarization-entangled photon-pair sources S1 and S2 for the experimental demonstration of the generation of the four-photon GHZ state using warm atomic ensembles of $^{87}$Rb atoms. The SFWM photon-pair sources are independently operated to investigate the properties of both sources toward realizing high-quality entangled states. The temporal waveforms of the two polarization-entangled SFWM photon-pair sources are obtained by means of the cross-correlation function between the signal and idler photons in both sources. In our experiment, the spectral width of the signal and idler photons corresponded to the Doppler-broadening linewidth of 540 MHz. The Cauchy–Schwarz inequality factor was estimated to be 10000, which clearly indicates the quantum nature of the paired photons. The counting rates of the signal and idler photons were measured to each be ~450 kHz, and the coincidence counting rate of the photon pair was obtained to be ~18 kHz with the coincidence window of 2.5 ns. Here, we note that the main cause of the brightness of our SFWM source and the strong time correlation between the paired photons is the superradiant effect, which is due to the collective two-photon coherence in the Doppler-broadened atomic ensemble [21].

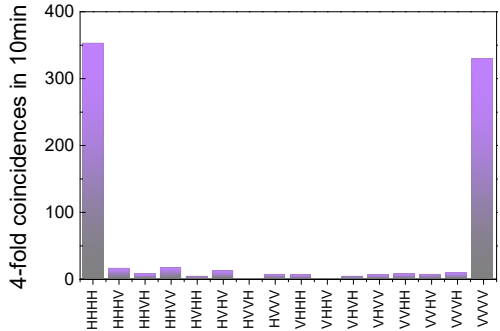

**Figure 2. Fourfold coincidence counts of the four-photon Greenberger–Horne–Zeilinger (GHZ) state from warm Rb ensembles.** Measured fourfold coincidence counts of 16 polarization combinations in the polarization-entangled GHZ state, corresponding to the $|G_4\rangle_{12'3'4}$ state defined by Eq. (2), from warm $^{87}$Rb atomic vapor.

Next, we discuss the generation of the four-photon GHZ state $|G_4\rangle_{12'3'4}$, corresponding to Eq. (2). To analyze the polarization-entangled $|G_4\rangle_{12'3'4}$ state, we measured the fourfold coincidence counts of all 16 possible polarization combinations according to the H/V basis of the four polarizers (P1, P2, P3, and P4). From Fig. 2, we can observe the fourfold events of both the HHHH and VVVV bases, corresponding to the desired $|G_4\rangle_{12'3'4}$ state. The fourfold coincidence in the HHHH basis was measured to be 350 events for the accumulation time of 600 s, corresponding to the fourfold coincidence count rate of 0.58/s. Although multiple photons are emitted from the atomic ensemble with a low pumping power of 0.15 mW and coupling power of 1.5 mW, our four-photon GHZ state is more than 10 times brighter than those reported previously in cold atoms [19].

We attempted to confirm that the $|G_4\rangle_{12'3'4}$ state are actually in coherent superposition. For this purpose, we performed HOM interference measurements (Fig. 3) via time-resolved coincidence at the PBS with signal photons 2 and 3 by using a time-tagging module for fourfold coincidence measurement. In Figs. 3(a) and (b), the x-axis indicates the arrival-time difference between the two independent CW-mode entangled photon pairs. For measuring the polarization-entangled state, we fixed the angles of polarizers P2 and P3 at 45°. As shown in Fig. 3(b), the HOM peak and dip fringes between the two independent heralded single-photons are observed with P1/P4 polarization orientations of 45°/45° and 45°/-45°, respectively. The HOM interference fringes shown in Fig. 3(b) were measured for a 60-min accumulation time with a 2.5-ns temporal window. The timing jitters of our single-photon detectors) SPDs were measured from ~0.4 ns, which is smaller than the temporal coherence time of the photons from the warm atomic ensemble. The raw visibilities were estimated to be 80.2 ± 0.7% and 79.6 ± 0.8%. Therefore, the results in Figs. 3(a) and 3(b) demonstrate the experimental realization of the four-photon polarization-entangled GHZ state from atomic ensembles.

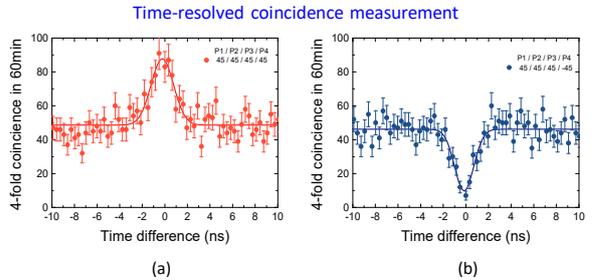

**Figure 3. Hong–Ou–Mandel (HOM) interference between two independent continuous wave (CW)-mode entangled photon pairs via a time-resolved coincidence measurement.** (a) HOM peak (red circles) and (b) HOM dip (blue circles) interference fringes of the polarization-entangled four-photon state generated from a warm atomic ensemble.

Finally, we performed the quantum-state tomography of the four-photon GHZ state to characterize the polarization entanglement of the four-photon GHZ state obtained from warm atomic ensembles. For the quantum-state tomography of the four-photon GHZ state, we measured the fourfold coincidence according to certain combinations of the QWP, HWP, and P components positioned in the paths of the four polarization-entangled photons (Fig. 1(b)). Figure 4 illustrates the real and imaginary parts of the reconstructed four-photon polarization density matrix $\hat{\rho}$ of the $|G_4\rangle_{1'2'3'4}$ state. The coincidence counting rates obtained from 256 measurement bases were used to reconstruct the density matrices for the four polarization-entangled states. The fidelity of the result in Fig. 4 to the ideal GHZ state was estimated to be $0.82 \pm 0.03$, based on raw data accumulated within 600 s.

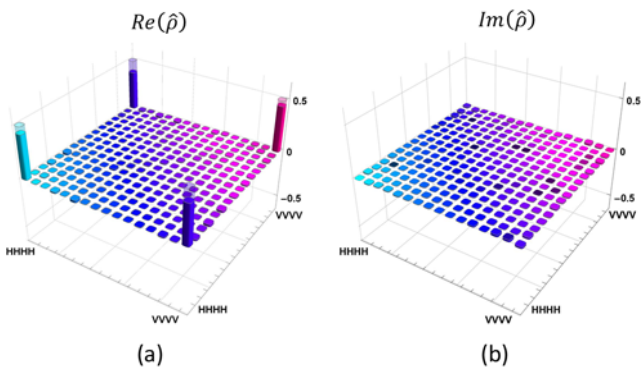

**Figure 4. Quantum-state tomography of the four-photon Greenberger–Horne–Zeilinger (GHZ) state from warm Rb ensembles.** (a) Real and (b) imaginary components of the density matrix reconstructed by using the maximum likelihood method for the GHZ state of the four polarization-entangled photons from a warm atomic ensemble.

In conclusion, we experimentally demonstrated the four-photon polarization-entangled GHZ state from warm atomic ensembles of $^{87}$Rb atoms for the first time. The average detection rate of the four-photon GHZ state was a relatively high value of 0.58 Hz owing to the collective two-photon coherence in the warm atomic ensemble. Moreover, the fidelity of the generated state was estimated to be $0.82 \pm 0.03$. Furthermore, we could confirm the generation of the polarization-entangled four-photon GHZ state by measuring the time-resolved HOM peak and dip fringes with raw visibilities of $80.2 \pm 0.7\%$ and $79.6 \pm 0.8\%$, respectively. We believe that our results can contribute to the realization of photonic quantum computing and optical quantum information processing based on atom–photon interactions.

**Acknowledgment**
This work was supported by the National Research Foundation of Korea (NRF) (2020M3E4A1080030) and the Ministry of Science and ICT (MSIT), Korea, under the Information Technology Research Center (ITRC) support program (IITP-2020-0-01606) supervised by the Institute of Information & Communications Technology Planning & Evaluation (IITP).

**Data availability**

The data that support the findings of this study are available from the corresponding author upon request.